\documentclass[preprintnumbers,amsmath,amssymbm,preprint]{revtex4}
\usepackage{epsfig}
\usepackage{graphicx}
\usepackage{amssymb}

\begin{document}

\title{Lower bound on the radii of light rings in traceless black-hole spacetimes}
\author{Shahar Hod}
\address{The Ruppin Academic Center, Emeq Hefer 40250, Israel}
\address{}
\address{The Hadassah Institute, Jerusalem 91010, Israel}
\date{\today}

\begin{abstract}
\ \ \ Photonspheres, curved hypersurfaces on which massless particles can perform closed geodesic motions around highly 
compact objects, are an integral part of generic black-hole spacetimes. 
In the present compact paper we prove, using analytical techniques, 
that the innermost light rings of spherically symmetric hairy black-hole
spacetimes whose external matter fields are characterized by a traceless energy-momentum tensor cannot 
be located arbitrarily close to the central black hole. 
In particular, we reveal the physically interesting fact that the non-linearly coupled Einstein-matter field equations set 
the lower bound $r_{\gamma}\geq {6\over5}r_{\text{H}}$ on the radii of traceless black-hole photonspheres, 
where $r_{\text{H}}$ is the radius of the outermost black-hole horizon. 
\end{abstract}
\bigskip
\maketitle

\section{Introduction}

Theoretical \cite{Bar,Chan,Shap,Hodub,Herne,Hodrec,YY} as well as observational \cite{Aki} studies have recently 
established the fact that closed light rings exist in the external spacetime regions of generic 
black holes. It has long been known that the presence of null circular geodesics 
in highly curved spacetimes has many implications 
on the physical and mathematical properties of the corresponding central black holes \cite{Bar,Chan,Shap,Herne,Hodns,Hodrec,YY,Aki,Mash,Goeb,Hod1,Dec,Hodhair,Hodfast,YP,
Hodm,Hodub,Lu1,Hodlwd,Pod,Ame,Ste}. 

For instance, the unstable circular motions of massless fields along closed null rings 
determine the characteristic relaxation timescale of a perturbed black-hole spacetime in the short wavelength (eikonal) 
regime \cite{Mash,Goeb,Hod1,Dec}. In addition, the optical appearance of a black hole to far away 
asymptotic observers is influenced 
by the presence of a light ring in the highly curved near-horizon region \cite{Pod,Ame,Ste}. 
Moreover, as measured by asymptotic observers, the equatorial null circular geodesic determines 
the shortest possible orbital period around a central non-vacuum 
black hole \cite{Hodfast,YP}.

Intriguingly, it has also been proved \cite{Hodhair,Hodub,Hodlwd,Hod1} that the innermost light ring of a 
non-trivial (non-vacuum) black-hole spacetime determines the non-linear spatial behavior of the supported hair. 
In particular, it has been revealed, using the non-linearly coupled Einstein-matter field equations, that the 
non-linear behavior of external hairy configurations which have a non-positive energy-momentum trace 
must extend beyond the null 
circular geodesic that characterizes the curved black-hole spacetime \cite{Hodhair,Hodub,Hodlwd,Hod1}.   

Motivated by the well established fact that null circular geodesics (closed light rings) 
are an important ingredient of generic black-hole spacetimes \cite{Bar,Chan,Shap,Hodub,Herne,Hodrec,YY,Aki}, 
in the present paper we raise the following physically intriguing question: 
How close can the innermost light ring of a central black hole be to its outer horizon? 

This is a seemingly simple question but, to the best of our knowledge, in the physics literature 
there is no general (model-independent) answer to it which is rigorously 
based on the Einstein equations.

In the present compact paper we shall reveal the fact 
that, for spherically symmetric hairy black-hole spacetimes whose supported field configurations are characterized by 
a traceless energy-momentum tensor, the non-linearly coupled Einstein-matter field equations provide 
an explicit quantitative answer to this physically important question. 
In particular, we shall explicitly prove 
that the radii of light rings in spherically symmetric traceless hairy black-hole
spacetimes are bounded from below by the functional relation 
\begin{equation}\label{Eq1}
r_{\gamma}\geq {6\over5}r_{\text{H}}\  ,
\end{equation}
where $r_{\text{H}}$ is the radius of the outermost horizon. 

It is worth noting that our theorem, to be presented below, is valid for 
the canonical family of colored black-hole spacetimes that characterize 
the non-linearly coupled Einstein-Yang-Mills (EYM) field theory (see \cite{Volk,Notesta} and references therein). 
In particular, it is worth emphasizing the fact that the highly non-linear character 
of the coupled Einstein-Yang-Mills field equations has restricted most former studies 
of this physically important field theory to the numerical regime. 
It is therefore of physical interest to reveal, using purely {\it analytical} techniques, 
some of the generic physical characteristics of this highly non-linear field theory. 
This is one of the main goals of the present paper.

\section{Description of the system}

We shall study, using analytical techniques, the radial locations of compact photonspheres (closed light rings) in spherically symmetric 
hairy black-hole spacetimes which are described by the curved line element
\cite{Hodfast,Hodm,Noteunit}
\begin{equation}\label{Eq2}
ds^2=-e^{-2\delta}\mu dt^2 +\mu^{-1}dr^2+r^2(d\theta^2 +\sin^2\theta d\phi^2)\  ,
\end{equation}
where $\{t,r,\theta,\phi\}$ are the Schwarzschild-like coordinates of the spacetime. 

The radial functional behaviors of the 
matter-dependent metric functions $\mu=\mu(r)$ and $\delta=\delta(r)$ are determined by 
the non-linearly coupled Einstein-matter field 
equations $G^{\mu}_{\nu}=8\pi T^{\mu}_{\nu}$ \cite{Hodfast,Hodm}:
\begin{equation}\label{Eq3}
{{d\mu}\over{dr}}=-8\pi r\rho+{{1-\mu}\over{r}}\
\end{equation}
and
\begin{equation}\label{Eq4}
{{d\delta}\over{dr}}=-{{4\pi r(\rho +p)}\over{\mu}}\  ,
\end{equation}
where the radially-dependent matter functions \cite{Bond1}
\begin{equation}\label{Eq5}
\rho\equiv-T^{t}_{t}\ \ \ \ ,\ \ \ \ p\equiv T^{r}_{r}\ \ \ \ , \ \ \ \ p_T\equiv T^{\theta}_{\theta}=T^{\phi}_{\phi}\
\end{equation}
in the differential equations (\ref{Eq3}) and (\ref{Eq4}) 
are respectively the energy density, the radial pressure, and the tangential pressure of 
the external matter configurations in the non-trivial (non-vacuum) black-hole spacetime (\ref{Eq2}). 

The radial metric functions $\{\mu,\delta\}$ of the black-hole spacetime are 
characterized by the horizon boundary relations \cite{Bekreg}
\begin{equation}\label{Eq6}
\mu(r=r_{\text{H}})=0\
\end{equation}
and
\begin{equation}\label{Eq7}
\delta(r=r_{\text{H}})<\infty\ \ \ \ ; \ \ \ \ [d\delta/dr]_{r=r_{\text{H}}}<\infty\  .
\end{equation}
In addition, the asymptotic functional relations \cite{Bekreg}
\begin{equation}\label{Eq8}
\mu(r\to\infty)\to1\ 
\end{equation}
and
\begin{equation}\label{Eq9}
\delta(r\to\infty)\to0\
\end{equation}
characterize the metric functions of asymptotically flat black-hole spacetimes. 

Our theorem, to be presented below, is based on the assumption that the external matter fields 
respect the dominant energy condition, which implies that the energy density 
is positive semi-definite \cite{Bekreg},
\begin{equation}\label{Eq10}
\rho\geq0\  ,
\end{equation}
and that it bounds from above the absolute values of the pressure components of the matter fields \cite{Bekreg}:
\begin{equation}\label{Eq11}
|p|, |p_T|\leq\rho\  .
\end{equation}
In addition, we shall assume that the external matter fields are characterized by a traceless 
energy-momentum tensor: 
\begin{equation}\label{Eq12}
T=0\  ,
\end{equation}
where $T=-\rho+p+2p_T$. In particular, the analytically derived lower bound on the characteristic 
radii of compact photonspheres [see Eq. (\ref{Eq31}) below] would be valid for the well-known 
colored black-hole spacetimes that characterize the composed Einstein-Yang-Mills field theory \cite{Volk}. 

Taking cognizance of the Einstein field equation (\ref{Eq3}), one finds the functional relation 
\begin{equation}\label{Eq13}
\mu(r)=1-{{2m(r)}\over{r}}\ 
\end{equation}
for the dimensionless metric function $\mu(r)$, where the radially-dependent physical parameter 
\begin{equation}\label{Eq14}
m(r)=m(r_{\text{H}})+\int_{r_{\text{H}}}^{r} 4\pi r^{2}\rho(r)dr\
\end{equation}
is the gravitational mass which is contained within an external sphere of radius $r\geq r_{\text{H}}$. 
Here $m(r_{\text{H}})$, which is characterized by the simple relation
\begin{equation}\label{Eq15}
m(r=r_{\text{H}})={{r_{\text{H}}}\over{2}}\  ,
\end{equation}
is the horizon mass (the mass contained within the black hole). 

\section{Lower bound on the radii of light rings in spherically symmetric traceless black-hole spacetimes}

In the present section we shall address the following question: How close 
can a black-hole photonsphere be to its outer horizon? 
Intriguingly, below we shall prove that an explicit answer to this physically important question, which is based 
on the non-linearly coupled Einstein-matter field equations, can be given for 
non-trivial (non-vacuum) hairy black-hole spacetimes whose external matter fields are characterized by 
a traceless energy-momentum tensor. In particular, we shall reveal the fact that 
the innermost light rings cannot be located arbitrarily close to 
the outer horizons of the central black holes. 

The radial locations of null circular geodesics (closed light rings) in spherically symmetric hairy 
black-hole spacetimes are determined by the roots of the dimensionless function \cite{Hodhair}
\begin{equation}\label{Eq16}
{\cal N}(r)\equiv 3\mu-1-8\pi r^2p\  .
\end{equation}
Taking cognizance of the fact that non-extremal black holes are 
characterized by the dimensionless horizon relations \cite{Bekreg}
\begin{equation}\label{Eq17}
0\leq 8\pi r^2_{\text{H}}\rho(r_{\text{H}})=-8\pi r^2_{\text{H}}p(r_{\text{H}})<1\  ,
\end{equation}
one finds that the function (\ref{Eq16}) is characterized by the horizon boundary condition [see Eq. (\ref{Eq6})] 
\begin{equation}\label{Eq18}
{\cal N}(r=r_{\text{H}})<0\  .
\end{equation}
In addition, from Eqs. (\ref{Eq8}), (\ref{Eq11}), (\ref{Eq13}), and (\ref{Eq14}) one deduces the asymptotic functional 
behavior 
\begin{equation}\label{Eq19}
r^2p\to0\ \ \ \ \text{for}\ \ \ \ r\to\infty\  ,
\end{equation}
which implies the simple radial behavior 
\begin{equation}\label{Eq20}
{\cal N}(r\to\infty)\to 2\  .
\end{equation}

The characteristic properties (\ref{Eq18}) and (\ref{Eq20}) of the dimensionless radial function (\ref{Eq16}) 
guarantee the existence of an external compact sphere with the 
property $r=r_{\gamma}>r_{\text{H}}$ for which
\begin{equation}\label{Eq21}
{\cal N}(r=r_{\gamma})=0\
\end{equation}
and
\begin{equation}\label{Eq22}
\Big[{{d{\cal N}}\over{dr}}\Big]_{r=r_{\gamma}}\geq0\  .
\end{equation}
The functional relations (\ref{Eq21}) and (\ref{Eq22}) determine the radial location of the 
innermost light ring which characterizes the spherically symmetric non-vacuum (hairy) 
black-hole spacetime (\ref{Eq2}).

Before proceeding, it is worth emphasizing that it has recently been proved \cite{Hodrec}, 
using the non-linearly coupled Einstein-matter field equations, that extermal black-hole spacetimes are characterized 
by the horizon relations ${\cal N}(r=r_{\text{H}})=0$ and $[{{d{\cal N}}/{dr}}]_{r=r_{\text{H}}}<0$ 
which, together with the asymptotic radial behavior ${\cal N}(r\to\infty)\to 2$ [see Eq. (\ref{Eq20})] of the 
dimensionless function (\ref{Eq16}), guarantee that extremal black holes, like non-extremal ones, 
possess external light rings (with $r=r_{\gamma}>r_{\text{H}}$) which are 
characterized by the functional properties (\ref{Eq21}) and (\ref{Eq22}). 
Thus, our analysis is also valid for spherically symmetric extremal black-hole spacetimes.   

Taking cognizance of the Einstein equations (\ref{Eq3}) and (\ref{Eq4}) together with the characteristic conservation equation
\begin{equation}\label{Eq23}
T^{\mu}_{r ;\mu}=0\  ,
\end{equation}
one finds the gradient relation
\begin{equation}\label{Eq24}
{{d}\over{dr}}(r^2p)={{r}\over{2\mu}}\Big[(3\mu-1-8\pi r^2p)(\rho+p)+2\mu(-\rho-p+2p_{T})\Big]\  ,
\end{equation}
which yields the functional relation [see Eqs. (\ref{Eq3}) and (\ref{Eq16})] \cite{Hodrole}
\begin{equation}\label{Eq25}
\Big[{{d{\cal N}}\over{dr}}\Big]_{r=r_{\gamma}}={{2}\over{r_{\gamma}}}\big[1-8\pi r^2_{\gamma}(\rho+p_T)\big]\  .
\end{equation}

Substituting Eq. (\ref{Eq25}) into (\ref{Eq22}) and using the trace relation (\ref{Eq12}) for 
the external matter fields, one obtains the relation
\begin{equation}\label{Eq26}
0\leq[1-8\pi r^2(\rho+p_T)]_{r=r_{\gamma}}=[1-12\pi r^2\rho+4\pi r^2p]_{r=r_{\gamma}}\
\end{equation}
which, using the dominant energy condition (\ref{Eq11}), yields the characteristic dimensionless inequality
\begin{equation}\label{Eq27}
0\leq [1+16\pi r^2p]_{r=r_{\gamma}}\
\end{equation}
at the radial location of the black-hole innermost photonsphere. 
Furthermore, substituting into (\ref{Eq27}) the relation (\ref{Eq21}), 
which characterizes the null circular geodesics of the black-hole spacetime (\ref{Eq2}), 
one obtains the inequality 
\begin{equation}\label{Eq28}
[6\mu(r)-1]_{r=r_{\gamma}}\geq0\
\end{equation}
which, using the functional relation (\ref{Eq13}), can be written in the form  
\begin{equation}\label{Eq29}
\Big[{{m(r)}\over{r}}\Big]_{r=r_{\gamma}}\leq{{5}\over{12}}\  .
\end{equation}
Finally, taking cognizance of Eqs. (\ref{Eq10}), (\ref{Eq14}), (\ref{Eq15}), and (\ref{Eq29}), 
one obtains the series of inequalities
\begin{equation}\label{Eq30}
r_{\gamma}\geq {{12}\over{5}}m(r_{\gamma})\geq {{12}\over{5}}m(r_{\text{H}})={{6}\over{5}}r_{\text{H}}\  .
\end{equation}

It is interesting to point out that the canonical family of electrically charged Reissner-Nordstr\"om black-hole 
spacetimes are characterized by the relations $r_{\text{H}}=M+(M^2-Q^2)^{1/2}$ 
and $r_{\gamma}={1\over2}[3M+(9M^2-8Q^2)^{1/2}]$ \cite{Chan,NoteMQ}, in which case one finds that the 
dimensionless ratio $r_{\gamma}/r_{\text{H}}$ is a monotonically increasing function of the dimensionless 
charge-to-mass ratio $|Q|/M$ of the black hole from the value $r_{\gamma}/r_{\text{H}}=3/2$ for $Q=0$ 
to the value $r_{\gamma}/r_{\text{H}}=2$ for the extremal black hole with $|Q|=M$. 
Thus, charged Reissner-Nordstr\"om black-hole spacetimes respect the analytically derived lower bound (\ref{Eq30}). 

\section{Summary}

The non-linearly coupled Einstein-matter field equations of general relativity predict 
the existence of compact photonspheres in the 
external regions of curved black-hole spacetimes. 
In particular, it is well established in the physics literature that closed 
light rings (null circular geodesics on which photons and gravitons 
can perform closed orbital motions around highly compact astrophysical objects) are of central importance 
in determining the physical, mathematical, and observational properties of generic (non-vacuum) 
black-hole spacetimes \cite{Bar,Chan,Shap,Herne,Hodns,Hodrec,YY,Aki,Mash,Goeb,Hod1,Dec,Hodhair,Hodfast,YP,Hodm,Hodub,Lu1,Hodlwd,Pod,Ame,Ste}. 

Motivated by the important roles that photonspheres play in the physics of black holes, 
in the present paper we have addressed the following question: How close 
can the black-hole innermost light ring be to the outer horizon of the corresponding central black hole? 
Perhaps somewhat surprisingly, to the best of our knowledge 
there is no general answer to this intriguing question in the physics literature. 

Interestingly, in the present compact paper we have proved, using analytical techniques, 
that an explicit answer to this physically important question can be given for spherically symmetric 
black-hole spacetimes whose external hairy configurations are characterized by 
a traceless energy-momentum tensor 
[It is worth noting that our main focus here is on the canonical family of colored black-hole spacetimes that 
characterize the non-linearly coupled Einstein-Yang-Mills field equations \cite{Volk}. 
However, it should be emphasized that our analytically derived results are also valid 
for any Einstein-matter field theory for which the external matter 
fields satisfy the traceless energy-momentum condition (\ref{Eq12})]. 

In particular, we have presented a remarkably compact theorem that reveals 
the physically interesting fact that the non-linearly coupled 
Einstein-matter field equations set the dimensionless lower bound [see Eq. (\ref{Eq30})] 
\begin{equation}\label{Eq31}
{{r_{\gamma}-r_{\text{H}}}\over{r_{\text{H}}}}\geq{{1}\over{5}}\
\end{equation}
on the radii of photonspheres (closed light rings) in spherically symmetric \cite{Notens} 
traceless hairy black-hole spacetimes.


\bigskip
\noindent {\bf ACKNOWLEDGMENTS}

This research is supported by the Carmel Science Foundation. I thank
Yael Oren, Arbel M. Ongo, Ayelet B. Lata, and Alona B. Tea for
stimulating discussions.


\begin{thebibliography}{99}
\bibitem{Notesta} It is worth emphasizing the fact that the EYM colored black holes are known to be unstable, see 
\cite{Volk} and references therein.
\bibitem{Noteunit} We shall use natural units in which $G=c=1$.
\bibitem{NoteMQ} Here $M$ and $Q$ are respectively the mass and electric charge of 
the Reissner-Nordstr\"om black hole.
\bibitem{Notens} Note that non-spherically symmetric spinning black holes may 
violate the lower bound (\ref{Eq31}) which is derived analytically in the present paper under the explicit assumption of spherical symmetry. In particular, co-rotating light rings of near-extremal (rapidly-spinning) Kerr black holes 
are characterized by the relation $({{r_{\gamma}-r_{\text{H}}})/{r_{\text{H}}}}\to0^+$ in the $a/M\to1^-$ limit 
(here $M$ and $J\equiv Ma$ are respectively the mass and angular momentum of 
the spinning Kerr black hole).
\end{thebibliography}
\end{document}